# $BiS_2$ - based superconductivity in F-substituted $NdOBiS_2$

Satoshi Demura,[1,2] Yoshikazu Mizuguchi,[1,3] Keita Deguchi,[1,2] Hiroyuki Okazaki,[1]

Hiroshi Hara,[1,2] Tohru Watanabe,[1,2] Saleem James Denholme,[1] Masaya Fujioka,[1]

Toshinori Ozaki,[1] Hiroshi Fujihisa,[4] Yoshito Gotoh,[4] Osuke Miura,[3] Takahide Yamaguchi,[1]

Hiroyuki Takeya,[1] and Yoshihiko Takano[1,2]

E-mail : Demura.Satoshi@nims.go.jp

[1]National Institute for Materials Science, 1-2-1 Sengen, Tsukuba, Ibaraki 305-0047, Japan

[2]University of Tsukuba, Graduate School of Pure and Applied Sciences, Tsukuba, Ibaraki 305-8577, Japan

[3]Tokyo Metropolitan University, Graduate School of Science and Engineering, Hachioji, Tokyo 192-0397, Japan

[4]National Institute of Advanced Industrial Science and Technology (AIST), Tsukuba Central 5, 1-1-1, Higashi, Tsukuba 305-8565, Japan.

**Abstract**

We have successfully synthesized a new $BiS_2$-based superconductor $NdOBiS_2$ with F-doping. This compound is composed of superconducting $BiS_2$ layers and blocking NdO layers, which indicates that the $BiS_2$ layer is the one of the common superconducting layers like the $CuO_2$ layer of cuprates or Fe-As layer of Fe-based superconductors. We can obtain $NdO_{1-x}F_xBiS_2$ with bulk superconductivity by a solid-state reaction under ambient pressure. Therefore, $NdO_{1-x}F_xBiS_2$ should be the suitable material to elucidate the mechanism of superconductivity in the $BiS_2$-layer.

Quite recently, Y. Mizuguchi et al. reported superconductivity in the novel $BiS_2$-based superconductor $Bi_4O_4S_3$ with a superconducting transition temperature ($T_c$) of 8.6 K [1]. This material has a layered structure composed of superconducting $BiS_2$ layers and blocking layers of $Bi_4O_4(SO_4)_{1-x}$, where $x$ indicates the defects of $SO_4^{2-}$ ions at the interlayer sites. The stacking structure of the superconducting and blocking layer is analogous to those of high-$T_c$ cuprates [2-5] and Fe-based superconductors [6-14]. In both systems, their $T_c$ can be enhanced by changing the blocking layers. In order to enhance the $T_c$ of the $BiS_2$-based superconductor, the investigation of exchanging the blocking layer will be of great interest.

Soon after the discovery of $Bi_4O_4S_3$, a new $BiS_2$-based superconductor $LaO_{1-x}F_xBiS_2$ was reported [15]. This compound consists of the same superconducting layer but with different blocking layers compared to that of $Bi_4O_4S_3$. Furthermore, the superconductivity shows a relatively high $T_c$ of 10.6 K. This fact suggests that the $BiS_2$ layer is the common superconducting layer, and the blocking layer contributes to the enhancement of the $T_c$ in this system. $LaO_{1-x}F_xBiS_2$ shows a small superconducting volume fraction for ambient pressure but achieves bulk superconductivity under high pressure. This result suggests that high pressure would promote F substitution. Therefore, chemical pressure for the exchange of La by Nd possibly induces the promotion of F substitution. Here, we report that a new $BiS_2$-based superconductor $NdO_{1-x}F_xBiS_2$ ($x$ = 0.1 - 0.7) can be obtained by a solid-state reaction under ambient pressure.

The polycrystalline samples of $NdO_{1-x}F_xBiS_2$ ($x$ = 0.1 - 0.7) were prepared by a solid-state reaction. Mixtures of Bi grains, $Bi_2S_3$ grains, $Nd_2O_3$ powders, $Nd_2S_3$ powders, and $NdF_3$ powders with nominal compositions of $NdO_{1-x}F_xBiS_2$ ($x$ = 0.1 - 0.7) were ground, pelletized, and sealed into an evacuated quartz tube. The tube was heated at 800 °C for 10 h. The obtained samples were characterized by X-ray diffraction with Cu-K$\alpha$ radiation using the $\theta$-$2\theta$ method. The temperature dependence of magnetization was measured by a superconducting quantum interface device (SQUID) magnetometer with an applied field of 1 Oe. The resistivity measurements were performed using the four-terminal method from 300 to 2 K. In order to investigate an upper critical field of $NdO_{1-x}F_xBiS_2$, the temperature dependence of resistivity between 10 and 2 K was measured under a magnetic field of up to 7 T.

Figure 1(a) shows the X-ray diffraction profile for the powdered samples of $NdO_{1-x}F_xBiS_2$ ($x$ = 0.1 - 0.7). With the exception of a few minor peaks relating to impurity phases, all of the peaks can be characterized to space group $P4/nmm$. The nominal $x$ dependence of the lattice constants $a$ and $c$ is summarized in Fig. 1(b) and (c). These lattice constants were estimated from $2\theta$ values and Miller indices. The $a$ lattice constant exhibits little change with increasing $x$ while the $c$ lattice constant dramatically decreases. The decrease of the lattice parameter $c$ indicates that F substitutes O since the ionic radius of F is smaller than that of O.

Figure 2(a) shows the temperature dependence of magnetic susceptibility for $NdO_{1-x}F_xBiS_2$ ($x$ = 0.1 - 0.7). A superconducting transition is observed for all samples. These samples with $x$ = 0.1 - 0.6 exhibit a high shielding volume fraction exceeding 90 % (2 K, ZFC), indicating the appearance of bulk superconductivity in those samples. The $x$ dependence of $T_c$ is plotted in Fig. 2(b). The $T_c$ varies like a bell curve with increasing $x$. The $NdO_{0.7}F_{0.3}BiS_2$ sample exhibits the optimal $T_c$ of all samples.

The temperature dependence of the resistivity for $NdO_{0.7}F_{0.3}BiS_2$ is shown in Fig. 3. Resistivity slightly decreases between 300 and 130 K. Below 130 K, resistivity gradually increases and the superconducting transition appears around 6 K. This behavior, where the resistivity increases with decreasing temperature is also observed in $LaO_{1-x}F_xBiS_2$. The onset and zero-resistivity temperatures are estimated to be $T_c^{onset}$ = 5.6 K and $T_c^{zero}$ = 5.2 K, respectively. Figure 4(a) shows the temperature dependence of the resistivity from 10 to 2 K under a magnetic field. The $T_c$ of $NdO_{0.7}F_{0.3}BiS_2$ decreases with increasing magnetic field. The upper critical field ($B_{c2}$) and the irreversibility field ($B_{irr}$) are plotted in Fig. 4(b). The $B_{c2}(0)$ was estimated to be 5.2 T with the data points at 0.4 ~ 2.0 T using the WHH theory, which gives $B_{c2}(0) = -0.69T_c(\mathrm{d}\,B_{c2}/\mathrm{d}T)|_{Tc}$ [16].

In conclusion, the $BiS_2$-based superconductor $NdO_{1-x}F_xBiS_2$ ($x$ = 0.1 - 0.7) has been successfully synthesized by a solid-state reaction method. Chemical pressure promotes the F substitution, which originated from the exchange of La by Nd. As a result, $NdO_{1-x}F_xBiS_2$ with bulk superconductivity can be obtained under ambient pressure. These results demonstrate that the $BiS_2$ layer is the common superconducting layer. Thus, if we synthesize materials with the stacking structure consisting of the $BiS_2$ layer and other blocking layers, new superconductors with the $BiS_2$-layer would be discovered.

## Acknowledgements

This work was partly supported by a Grant-in-Aid for Scientific Research from the Ministry of Education, Culture, Sports, Science and Technology (KAKENHI).

References

[1] Y. Mizuguchi et al., arXiv:1207.3145.

[2] J. G. Bednorz and K. Müller, *Z. Physik B Condensed Matter* **64**, 189-193 (1986).

[3] M. K. Wu, et al., *Phys. Rev. Lett.* **58**, 908–910 (1987).

[4] H. Maeda et al., *Jpn. J. Appl. Phys.* **27**, L209-L210 (1988).

[5] A.Schilling et al., *Nature* **363**, 56 - 58 (1993).

[6] Y. Kamihara et al., *J. Am. Chem. Soc.* **130**, 3296–3297 (2008).

[7] X. H.Chen et al., *Nature* **453**, 761-762 (2008).

[8] Z. A. Ren et al., Chinese Phys. Lett. 25, 2215 (2008).

[9] M. Rotter, M. Tegel and D. Johrendt, *Phys. Rev. Lett.* **101**, 107006(1-4) (2008).

[10] X. C. Wang et al., *Solid State Commun.* **148**, 538–540 (2008).

[11] F. C. Hsu et al., **105**, 14262–14264 (2008).

[12] K. W. Yeh et al., *EPL* **84**, 37002(p1-4) (2008).

[13] Y. Mizuguchi et al. *Appl. Phys. Lett.* **94**, 012503(1-3) (2009).

[14] J. Guo et al., *Phys. Rev. B* **82**, 180520(1-4) (2010).

[15] Y. Mizuguchi et al., arXiv:1207.3567

[16] N. R. Werthamer, E. Helfand, and P. C. Hohemberg, *Phys. Rev.* **147**, 295-302 (1966).

**Figure caption**

Fig. 1

(a) X-ray diffraction patterns of $NdO_{1-x}F_xBiS_2$ ($x$ = 0.1 - 0.7). Filled circles indicate the peaks of the impurity phases. (b) and (c) show the nominal $x$ dependence of the lattice constants $a$ and $c$ , respectively.

Fig. 2

(a) The temperature dependence of the magnetic susceptibility for $NdO_{1-x}F_xBiS_2$ ($x$ = 0.1 - 0.7).

(b) The F concentration dependence of the superconducting transition temperature for $NdO_{1-x}F_xBiS_2$ ($x$ = 0.1 - 0.7).

Fig. 3

The temperature dependence of resistivity for $NdO_{0.7}F_{0.3}BiS_2$ between 300 and 2 K.

Fig. 4

(a) The temperature dependence of the resistivity from 10 to 2 K under magnetic fields

(b) Magnetic field – temperature phase diagram for $NdO_{0.7}F_{0.3}BiS_2$. The dashed lines are liner fits to the data.

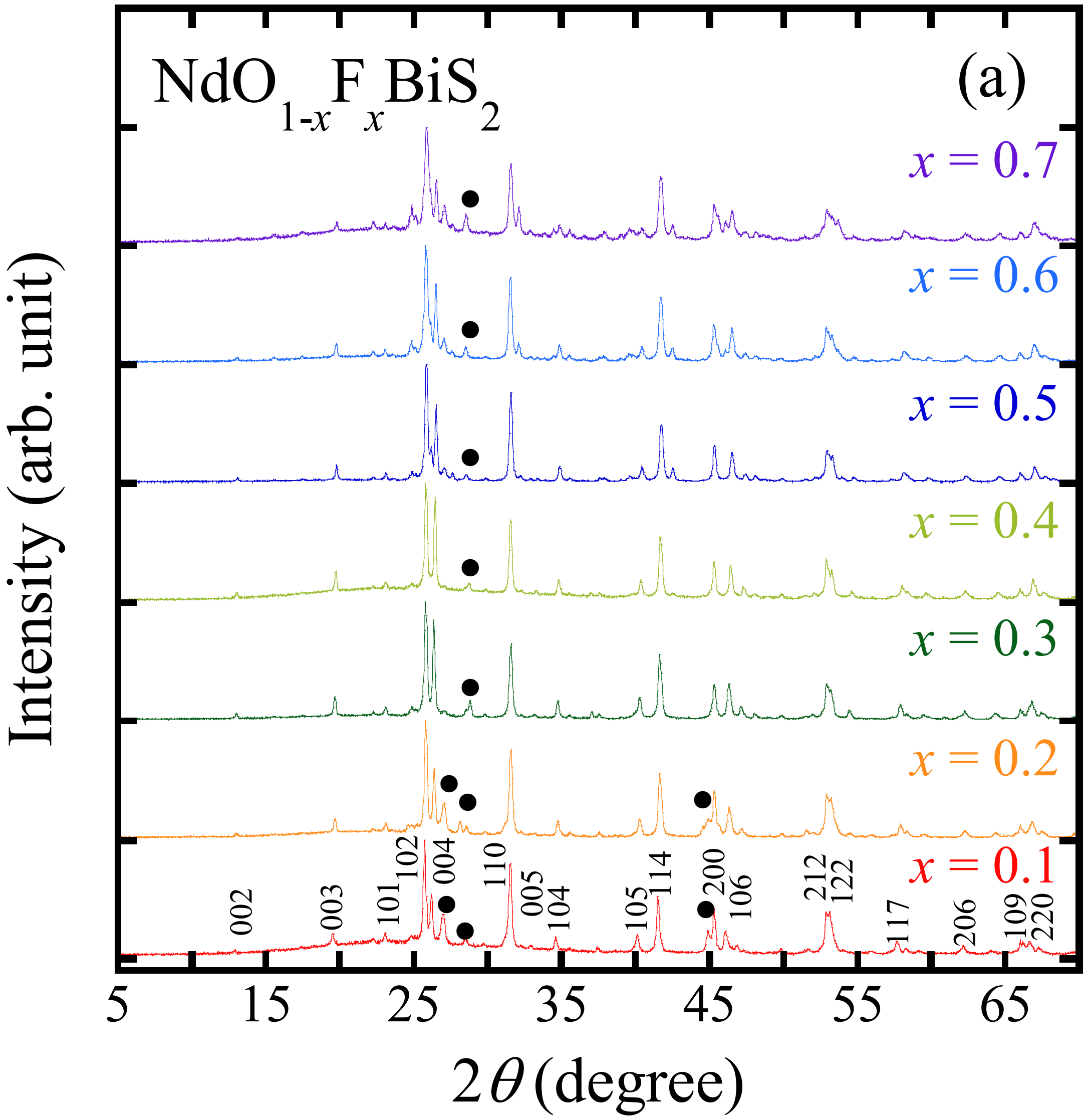

Fig. 1(a). S. Demura

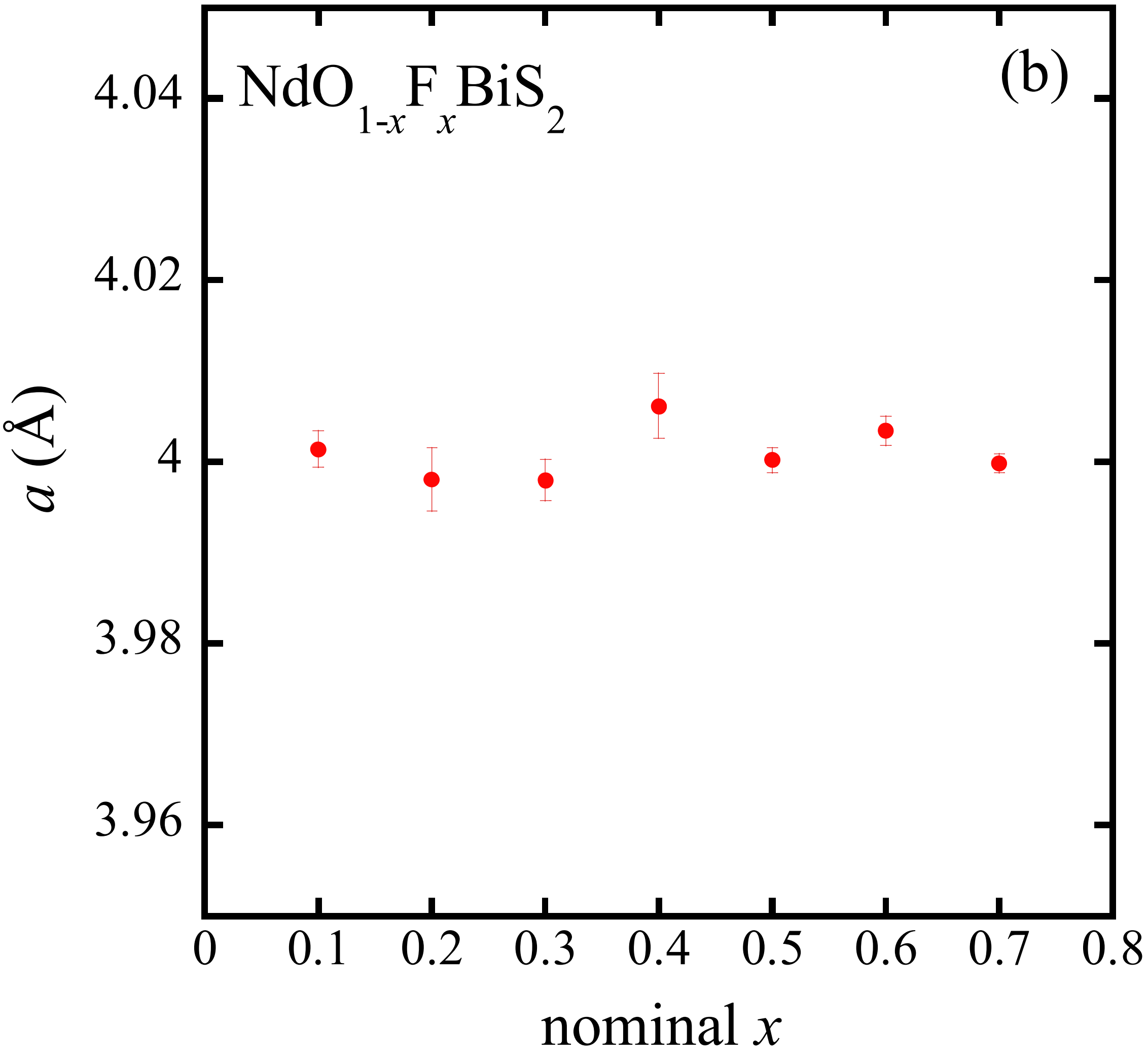

Fig. 1(b). S. Demura

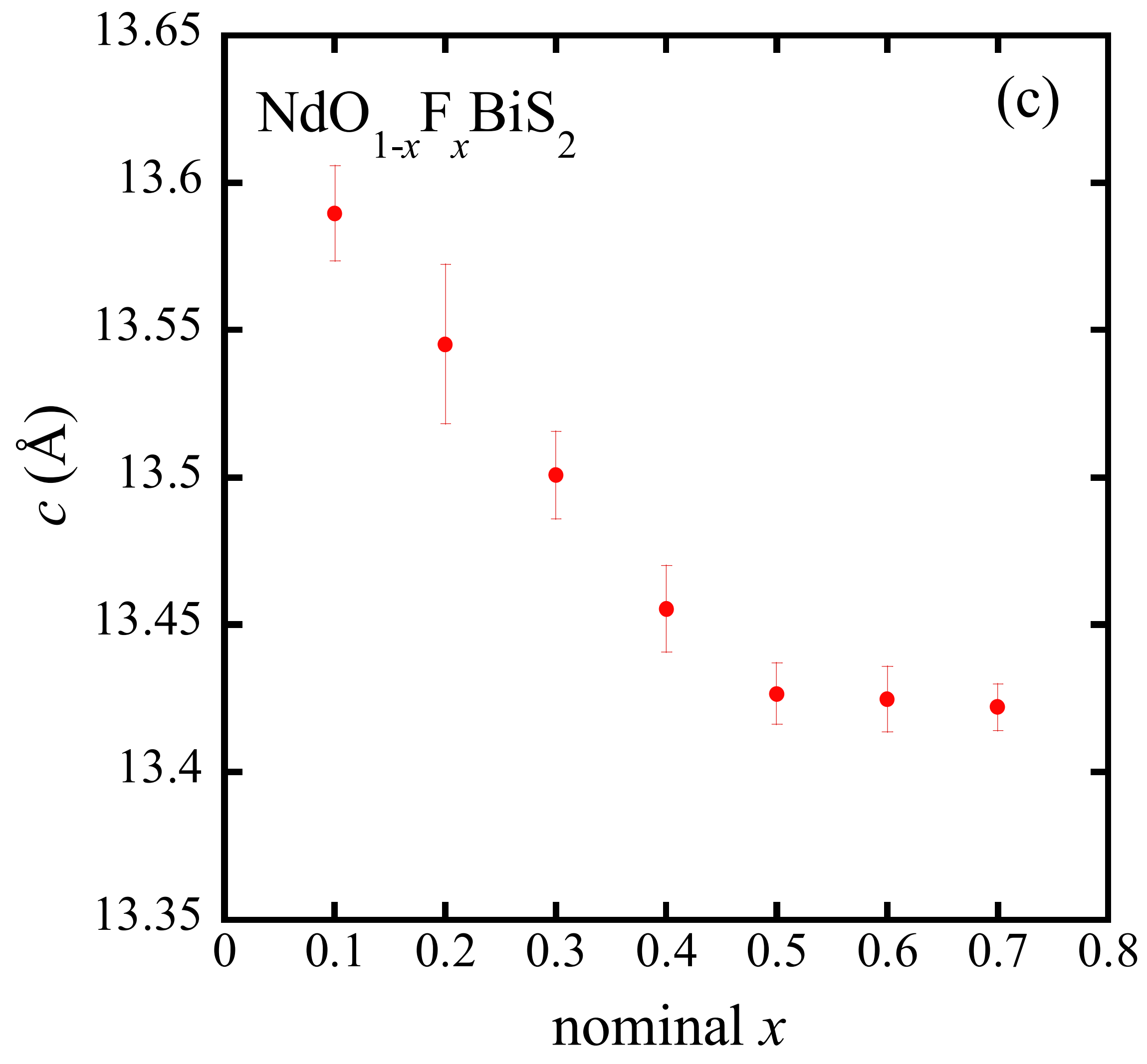

Fig. 1(c). S. Demura

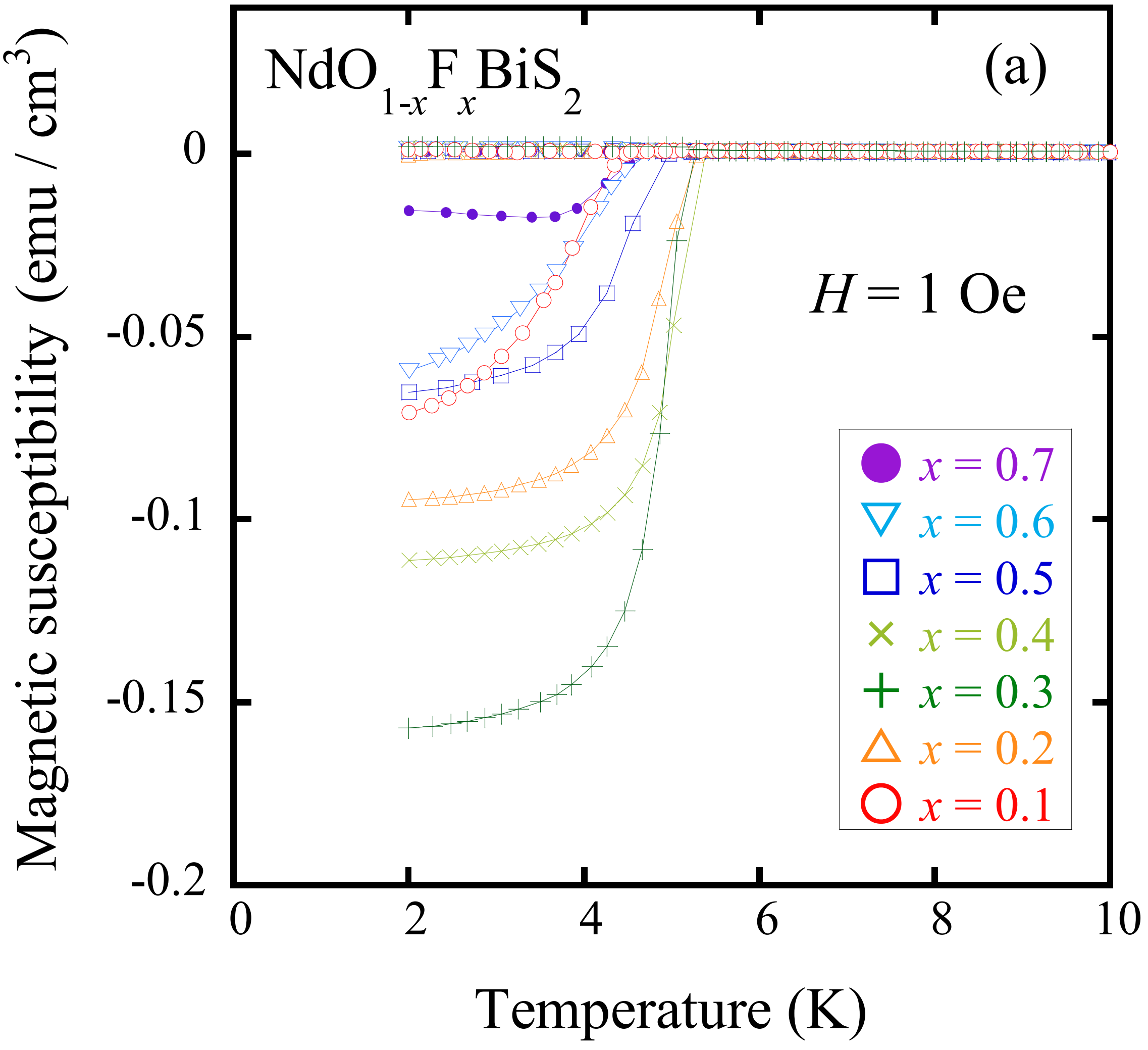

Fig. 2(a) S. Demura

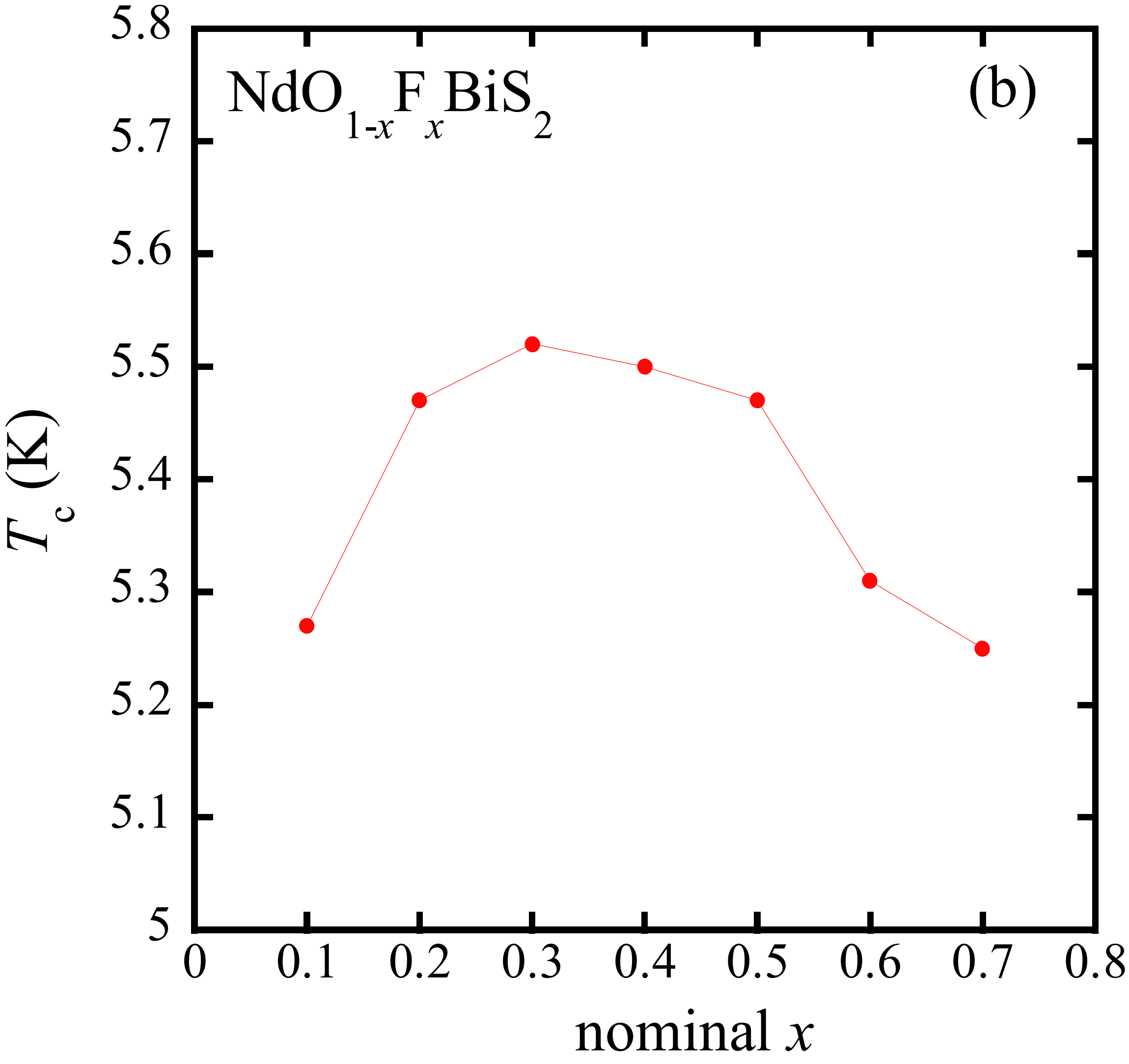

Fig. 2(b) S.Demura

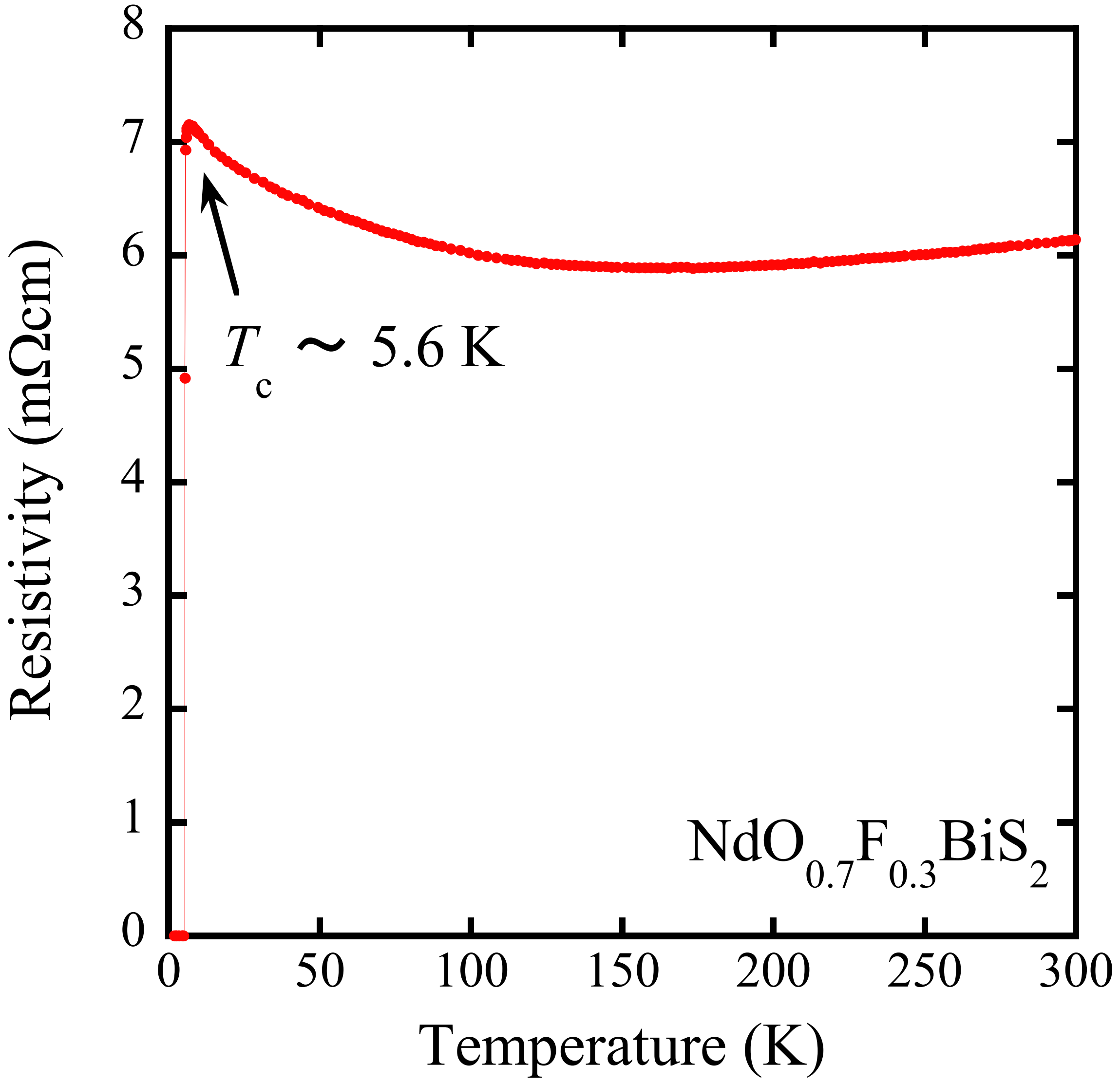

Fig. 3 S. Demura

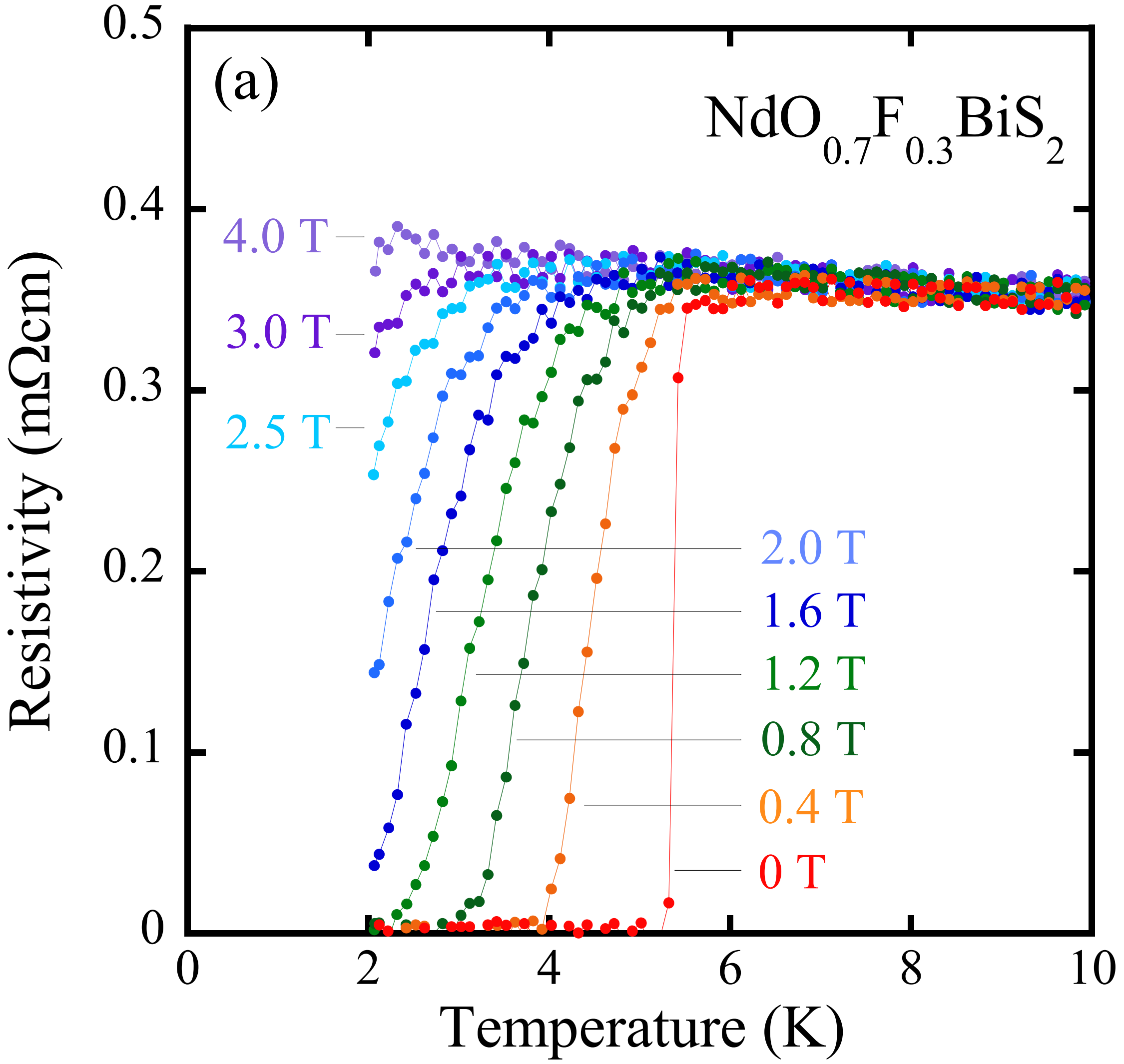

Fig. 4 S. Demura

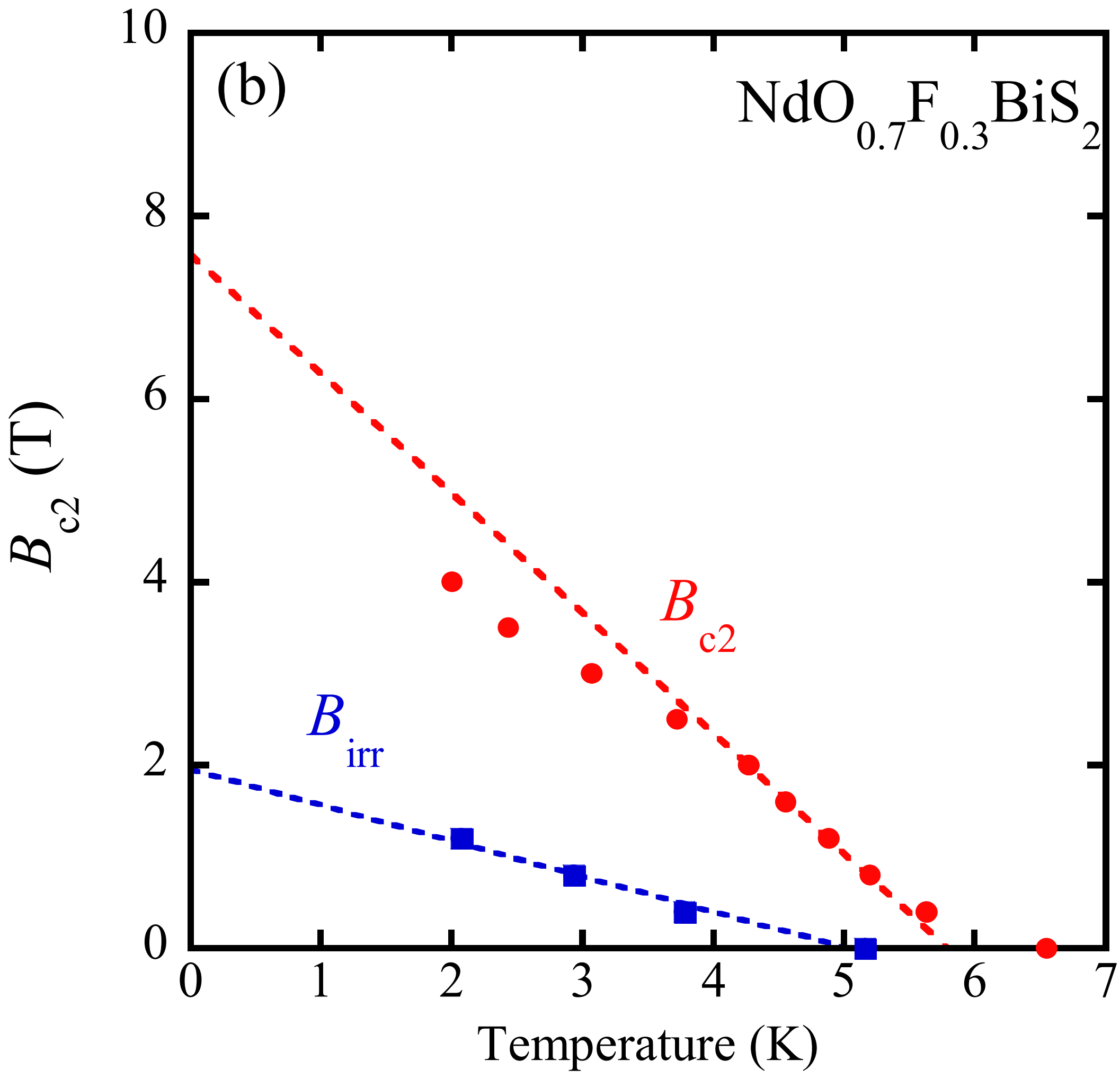

Fig. 4(b) S. Demura